\pdfoutput=1 
\documentclass[conference]{IEEEtran}
\IEEEoverridecommandlockouts
\usepackage{cite}
\usepackage{amsmath,amssymb,amsfonts}
\usepackage{graphicx}
\usepackage{textcomp}
\usepackage{xcolor}
\def\BibTeX{{\rm B\kern-.05em{\sc i\kern-.025em b}\kern-.08em
    T\kern-.1667em\lower.7ex\hbox{E}\kern-.125emX}}

\usepackage{algpseudocode}
\usepackage{xcolor}
\usepackage{soul}
\usepackage{libertine}
\usepackage{amsmath}
\usepackage[linesnumbered,ruled]{algorithm2e}
\usepackage{url}
\usepackage{mathtools}

\begin{document}

\title{A Graph-Based Customizable Handover Framework for LEO Satellite Networks
% A Graph-Based Framework For Customizable Handovers in LEO Satellite Networks
% A Customizable Handover Planning Framework for LEO Satellite Communication Networks \\
% {\footnotesize \textsuperscript{*}Note: Sub-titles are not captured in Xplore and
% should not be used}
% \thanks{.}
}

\author{
\IEEEauthorblockN{Mohamed Hozayen, Tasneem Darwish, Gunes Karabulut Kurt\IEEEauthorrefmark{5}, Halim Yanikomeroglu}
\IEEEauthorblockA{Department of Systems and Computer Engineering, Carleton University, Ottawa, ON, Canada \\
\IEEEauthorblockA{\IEEEauthorrefmark{5}Poly-Grames Research Center, Department of Electrical Engineering, Polytechnique Montréal, Montréal, QC, Canada} \\
Emails: \{mohamedhozayen, tasneemdarwish, guneskurt, halim\}@sce.carleton.ca\\
}}

\maketitle
% page numbering
% \thispagestyle{plain}
% \pagestyle{plain}

\begin{abstract}
 Future satellite networks are expected to have thousands of low Earth orbit (LEO) satellites orbiting Earth at very high speeds. User equipment (UE) communicating directly with LEO satellites will experience frequent handovers. Managing the handover process is complicated due to the high frequency of handovers and the availability of multiple LEO satellites as handover targets. In addition, as the status of the communication link between a UE and an LEO satellite varies in accordance with the visibility period of the satellite, initiating handovers at the right time will significantly affect the quality of service (QoS) of the communication provided. To address this problem, this work proposes a graph-based customizable handover framework that considers both the handover timing and target while selecting a handover sequence that maintains QoS. A time-based graph is designed where the vertices represent the satellites' instances over a certain period of time and the edges' weights are the customizable handover criteria (i.e., data rate and delay in this work). The appropriate sequence and timing of handovers that fulfill the required QoS are obtained by finding the shortest path in the graph. Discussion and simulations, which were conducted on the Starlink Phase I constellation, show the low complexity and performance advantages of the proposed handover framework. 
 \newline

\end{abstract}


\section{Introduction}
In future integrated vertical heterogeneous networks (VHetNets) \cite{alzenad2019coverage}, low Earth orbit (LEO) satellites will play a significant role in providing global communication and Internet services. Many commercial companies, including SpaceX and Amazon are launching LEO satellite constellations to provide global connectivity \cite{starlink}. Due to the low altitude of LEO satellites (500 km - 1,500 km), their propagation delays are lower than geostationary (GEO) satellites \cite{graph-based-handover}. Moreover, their low altitude reduces the amount of transmission power required and diminishes signaling attenuation, both of which are necessary to enable direct communication with user equipment (UE). 
Unfortunately, the deployment of LEO satellites has some disadvantages. 
Due to their high speeds, UE will need to handover its connection from one satellite to another every 5–10 minutes. In the near future, satellite mega-constellations will be able to provide communication services anywhere and anytime. However, the availability of several satellites over a certain geographic area makes handover decisions (i.e., choosing the handover timing and target satellite) a more complicated process. In addition, multiple applications with different quality of service (QoS) requirements will use satellite networks. Thus, to satisfy the needs of different applications, a customizable handover timing and target calculation is required. 

Some work has already been done to tackle handover issues for LEO satellite networks.
In \cite{1.2001-modeling}, the authors developed a hard handover scheme and a hybrid channel adaptive handover scheme by considering satellite signal strength relative to the elevation angle. 
The work in \cite{2.2003-evaluation} proposed a guaranteed handover procedure in non-geostationary satellite constellations requiring mutual visibility based on various criteria, including visible time, satellite capacity, elevation angle, and a combination of the same. 
The authors in \cite{4.2016-sealess} proposed a handover protocol based on a software-defined satellite network architecture, where the performance was evaluated on the basis of latency, throughput, and quality of experience for users.
The work in \cite{graph-based-handover} presented a graph-based framework to support handover decision strategies in LEO satellite networks on the basis of service time, elevation angle, and number of free channels.
In \cite{forecast-handover}, the authors proposed a multi-layer handover management framework, which introduced several handover procedures where the performance was evaluated on the basis of dropping probability and throughput.

Although several studies have been done on LEO satellite handover decisions, to the best of our knowledge, the joint consideration of satellite handover timing and targeting in planning the UE handover sequence has not yet been studied. In this work, we investigate the optimal sequence for handover timing and targeting that would fulfill user QoS requirements while considering parameters that would affect the handovers.

In so doing, we present a graph-based customizable handover planning framework for LEO satellite networks. To do this, we formulate the problem as a time-based graph by representing a satellite as a series of instances that reflect its position and link status over time. Satellite instances are represented as graph nodes, and each edge has a weighted sum value calculated using a customized set of handover parameters (i.e., data rate and delay in this work). The favorable sequence of handover timing and targets is determined by finding the shortest path in the graph. The framework is customizable, modular, flexible, and thus forward compatible with future integrated VHetNets. As the framework predicts future handover timing and targets, it will also be able to support soft handovers (connect before disconnect) leading to lower packet loss.

\section{Graph-Based Customizable Handover Planning Framework for LEO Satellite Communication Networks}

\subsection{Network Model}
\begin{figure}[htbp]
\centerline{\includegraphics[width=\linewidth]{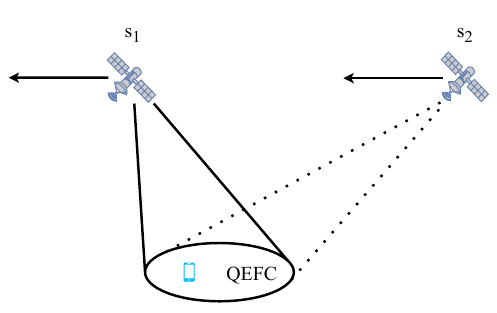}}
\caption{A quasi earth-fixed cell served by multiple satellites.}
\label{fig:scenario}
\end{figure}

\begin{figure}[htbp]
\centerline{\includegraphics[width=\linewidth]{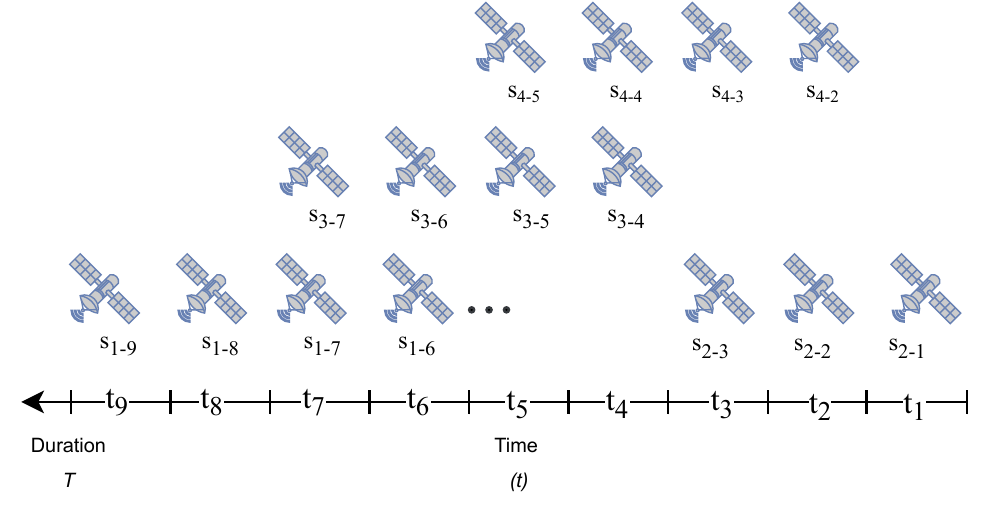}}
\caption{LEO satellite positions at different time instances.}
\label{fig:instances}
\label{fig:sat-timeline}
\end{figure}

In this work, we consider the scenario of a group of LEO satellites serving a quasi earth-fixed cell (QEFC)
% a fixed tracking area (TA) 
on Earth. This was proposed by 3GPP in Release 17 as a potential solution for non-geostationary satellite orbit (NGSO) mobility management \cite{3gpp-23.737-h00-Rel17}. As shown in Fig. \ref{fig:scenario}, when satellite $s_1$ approaches the minimum elevation angle, users in the QEFC need to perform a handover to satellite $s_2$ to maintain their current connections. 
In future LEO satellite mega-constellations, a single QEFC can be served by several satellites, which will create multiple targeted satellite options in the handover process. Due to satellite movement and the satellite's signal blocking in some areas (e.g., urban areas with high buildings), the quality of satellite to UE communication links varies in accordance with the satellite's visibility period. In addition, users or applications with different QoS requirements may necessitate customized handover decision-making. Thus, joint consideration of the handover timing and target (i.e., next satellite) is necessary to make optimized handover decisions.  
Fig. \ref{fig:instances} shows different instances of satellites $s_1$,  $s_2$, $s_3$, and $s_4$ over a time duration, $T$, divided into several short, equally sized time durations  ${t_{1},~t_{2},...,t_{i}}$.
The optimal time for the handover from satellite $s_1$ to any of the target satellites ($s_2$, $s_3$, or $s_4$) is affected by several criteria (e.g., satellites positions, link quality, required QoS), which are variable through time. Therefore, in the proposed framework we represent every satellite as several satellite instances to reflect changes in satellite communication status over time.
Accordingly, the satellite instance $s_{2-2}$ represents satellite $s_2$ at time duration $t_2$. Since future satellite networks will have thousands of LEO satellites, the coverage periods of multiple satellites will be overlapping.
We assume user positions and LEO satellite information are readily obtainable, e.g., by using the GPS-based approach as in \cite{leo-visibility-predict} and the simplified general perturbations (SGP4) model as in \cite{SGP4}.

\subsection{Graph-Based Problem Formulation for Handover Timing and Target}

Let us suppose that for a given user in a QEFC, to maintain a connection with certain QoS requirements, a sequence of handovers is required among satellites for a time duration, $T$. As shown in Fig. \ref{fig:instances}, we split $T$ into $n$ equal sized time durations, $t_i$, i.e., $\sum_{i=1}^{n}{t_i}=T$ and $t_i=(\mu_i \pm \lambda)$, where $\mu_i$ is the mean value of $t_i$ and $\lambda$ is the relaxation period. The satellites covering a certain QEFC during $T$ are mapped to an instance at every $t_i$ creating several instances of each satellite (e.g., $s_{2-1}, s_{2-2}, s_{2-3}$). The  satellite instances form a set $S=\{s_{1-1},s_{1-2},...,s_{1-n},s_{2-1},...,s_{k-n}\}$, where $k$ is the total number of satellites, $n$ is the total number of time instances, $n \times k$ is the total number of satellite instances, and each element $s_{j-i} \in S$ has a corresponding start and end periods, i.e., $\mu_i-\lambda$ and $\mu_i+\lambda$, respectively.
By considering satellite instances as graph nodes and the handover decision-making criteria as directed weighted edges, the sequence of handover timings and targets for period $T$ is determined by finding the highest ranking path in the directed weighted graph.

\begin{figure}[htbp]
\centerline{\includegraphics[width=\linewidth]{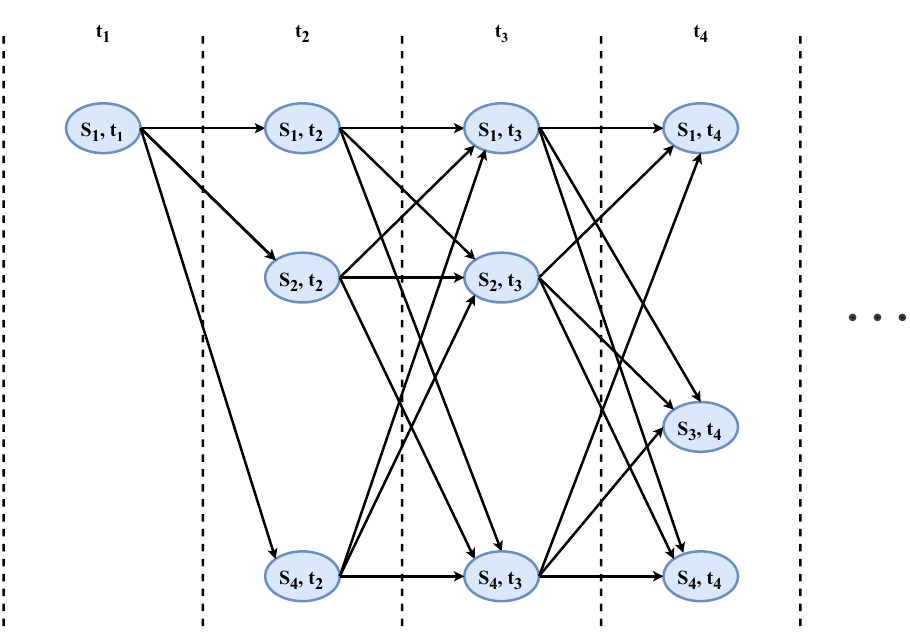}}
\caption{The directed weighted handover graph corresponding to Fig. \ref{fig:sat-timeline}.}
\label{fig:graph-example}
\end{figure}

Fig. \ref{fig:graph-example} shows the directed weighted time-based graph of the LEO satellites corresponding to Fig. \ref{fig:sat-timeline}. In the directed graph, the straight edges represent the transition between different instances of the same satellite (no handover is performed), whereas diagonal edges represent handovers between instances of two different satellites with an overlapping coverage period. Each directed edge weight in the graph is calculated as a weighted summation of pre-defined utility functions, which collectively form the handover decision-making criteria. The weighted summation model (WSM) is one of the multi-criteria decision making (MCDM) methods. Equation (\ref{eq:weighted-sum}) formulates the directed edge weight calculation, in general, as 

\begin{equation}
    W^{s_{j-i}} = \sum_{m=0}^{M}{w_{m}~U_{m}^{s_{j-i}}} 
    ,
    \label{eq:weighted-sum}
\end{equation}

\noindent where $W^{s_{j-i}}$ is the edge weight directed to the satellite instance $s_{j-i}$, $U_{m}^{s_{j-i}}$ is a utility function representing a handover criterion corresponding to satellite $s_{j}$ at time $t_{i}$, $w_m$ is a weighting factor of utility function $U_{m}^{s_{j-i}}$, and $M$ is the total number of considered utility functions in the handover decision-making, where $\sum_{m=0}^{M}{w_{m}=1}$. The handover framework can be customized to the requirements of each user by changing the weighting factors to give preference to certain utility functions and/or by using different utility functions (e.g., data rate, delay, and jitter). It should be noted that since the model is a linear combination of different utility functions, each utility function shall be normalized before calculating edge weights. This ensures that the model does not overlook a certain utility function over others due to its nature of high values. Nonetheless, the selection of which normalization technique to use is not the focus of this work. For example, delay is measured in milliseconds in satellite networks, whereas data rate is measured in megabits per seconds. The preference for a certain utility function is given by the weight $w_m$.

In this work, the data rate and delay are the criteria considered in the handover decision-making process. Thus, the generic equation (\ref{eq:weighted-sum}) can be specified as follows:

\begin{equation}
    W^{s_{j-i}} = 
    w_{d}~{U}_{d}^{s_{j-i}} + 
    w_{r}~{U}_{r}^{s_{j-i}}
    .
    \label{eq:weighted-sum-rate-delay}
\end{equation}

\noindent The favorable sequence of handovers in the graph is determined using the Dijkstra’s Algorithm to find the shortest path between two selected nodes, where the shortest path is the path with the lowest summation of weight edges. Therefore, lower delays are preferred, and therefore, 
\begin{equation}
{U}_{d}^{s_{j-i}} = \hat{PD}
,
\end{equation}
\noindent where $\hat{PD}$ is the normalized propagation delay. However, higher data rate values are preferred, and therefore, 
\begin{equation}
    {U}_{r}^{s_{j-i}} = 1 - \hat{R}
    ,
\end{equation}
\noindent where $\hat{R}$ is the normalized data rate.

\subsection{Customized Handover Sequence Selection Process}
The graph-based customizable handover framework selects handover timings and sequence of targeted satellites for the next time duration $T$. The selection process is detailed as follows:

\begin{enumerate}
    \item Select QEFC's and UE's location.
    \item Identify UE's QoS preferences and utility functions.
    \item Select preferred $T$ and $\lambda$.
    \item Obtain LEO satellite information accordingly.
    \item Choose appropriate channel model.
    \item Build UE's customized database table.
    \begin{enumerate}
        \item Row indexes are satellite numbers.
        \item Columns are time instances.
        \item Each cell represents a satellite instance of set $S$ that matches its corresponding column and row index. 
        \item Cells contain the calculated weighted sum of utility functions of the same satellite instance.
    \end{enumerate}
    \item Build a time-based graph as in Fig. \ref{fig:graph-example}.
        \begin{enumerate}
            \item The directed edge to a satellite instance is its own cell value.
            \item Add a virtual beginning node directed to the first instance of all satellites, i.e., $s_{j-1}$.
            \item Add a virtual end node for the last instance of all satellites, i.e, $s_{j-n}$, directs to.
        \end{enumerate} 
    \item Solve the graph from the virtual beginning node to the virtual end node using Dijkstra’s Algorithm.

\end{enumerate}

The solution is a sequence of nodes in the form of $s_{j-i}$, where $j$ is the satellite number and $i$ is the handover timing indication where $t_i$=($\mu_i \pm \lambda$). While $\lambda$, the relaxation period, provides some flexibility as to the exact time in which the handover can take place, in this paper the handover timing is taken at ($\mu_i + \lambda$).

\subsection{Channel Model}
The channel model expression mainly consists of path loss, Rician small-scale fading, and atmospheric fading \cite{channel-model-2}. The atmospheric fading, $A(d)$, is given by

\begin{equation}
    A(d)=10^{\frac{3d\chi}{10h}}
    ,
\end{equation}

\noindent where $\chi$ is the attenuation through the clouds and rain in dB/km, $d$ is the propagation distance between satellites and the UE, calculated by

\begin{equation}
    d=\sqrt{h^2+(x-o_{x})^2+(y-o_{y})^2}
    ,
\end{equation}

\noindent where $(o_{x},o_{y})$ is the position right below the satellite. The channel model expression is 

\begin{equation}
    G = (\frac{c_{light}}{4\pi d f_c})^2~A(d)~\varphi
    ,
\end{equation}

\noindent where $\varphi$ is the Rician small-scale fading. $c_{light}$ and $f_c$ are the speed of light and the carrier frequency, respectively. The received power is given by 

\begin{equation}
    P_{rx}=P_{tx} ~ G_{tx} ~ G ~ G_{rx}
    ,
\end{equation}

\noindent where $P_{tx}$ is the transmit power, $G_{tx}$ and $G_{rx}$ represent the antenna gains of the transmitter and receiver, respectively. The user data rate is given by the Shannon capacity theorem
\begin{equation}
    R = B \log(1+\frac{P_{rx}}{P_N})
    ,
    \label{eq:rate-shannon}
\end{equation}
\noindent where $B$ is the channel bandwidth, and $P_N$ is the noise power. Propagation delay is given by 

\begin{equation}
    PD = \frac{d}{c_{light}}
    .
    \label{eq:prop-delay}
\end{equation}

\section{Framework Evaluation}
\subsection{Complexity Analysis}
The highest ranking path in the constructed time-based graph is determined using the Dijkstra’s Algorithm by finding the shortest path between two nodes. The Dijkstra's Algorithm has a complexity of $O(E+V \log V)$ when a Fibonacci heap is used, where $E$ is the number of edges and $V$ is the number of nodes, or, in the context of this framework, the total number of satellite instances \cite{dijkstra-fibonacci}. Let $n=E+V$, therefore, 
$O(n) < O(E+V \log V) < O(n^2)$. 

As T is split into smaller equally sized time durations, $t_i$, $\lambda \rightarrow 0$, then $V,E$ $\rightarrow \infty$, and thus, the complexity of the handover graph approaches infinity while the handover solution approaches optimality:

\begin{equation}
    \lim_{\lambda\to 0} [H(S,T,\lambda),O(E+V \log V)] = [v^*, \infty]
    \label{eq:complexity}
    .
\end{equation}

\begin{figure}[htbp]
\centerline{\includegraphics[width=\linewidth]{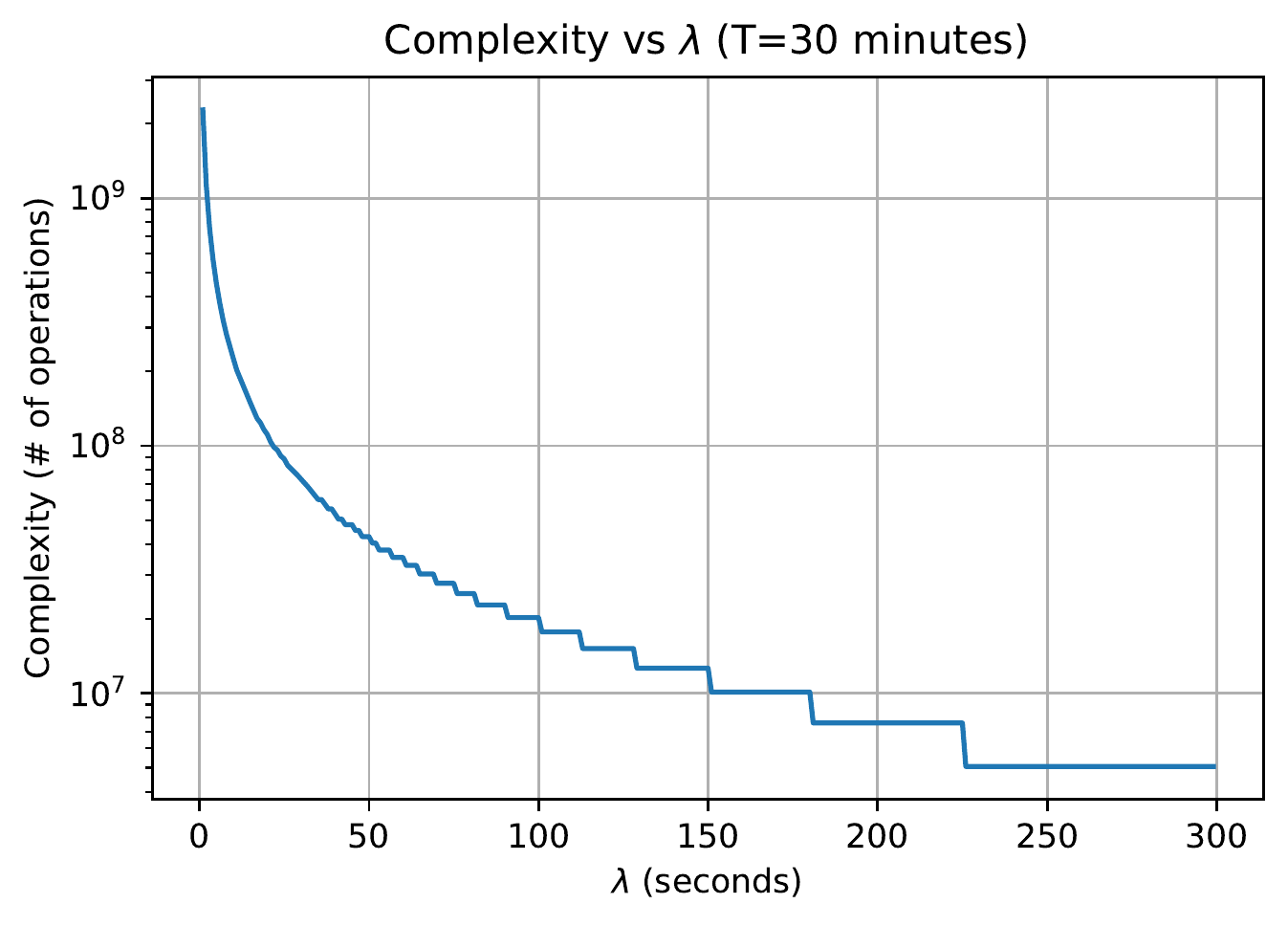}}
\caption{The complexity of finding the shortest path in the handover graph using all 1,584 satellites of Starlink Phase I Constellation using different $\lambda$ values.}
\label{fig:complexity-s-variable}
\end{figure}

Fig. \ref{fig:complexity-s-variable} shows the complexity analysis of finding the shortest path in the handover graph for different $\lambda$ values at $T=30$ minutes. In a worst case scenario, for a given QEFC, all 1,584 satellites of Starlink Phase I constellation would be used, but in reality far fewer satellites would be expected to fly over a QEFC during a 30-minutes period. To bring things into prospective, $1.8 \times 10^9$ operations are executed in half a second on a personal user device with a CPU frequency of $3.6$ GHz, given that one operation consumes one CPU cycle.

\subsection{Simulation Parameters}
The graph-based handover framework is simulated within a Starlink Phase I constellation environment generated by the well-known satellite constellation simulator STK version 12.1 \cite{stk}. The constellation consists of 1,584 satellites within 22 orbits (72 satellites per orbit) with an altitude of 550 km \cite{starlink}. The location of interest selected as the QEFC is Ottawa with one UE under evaluation. The downlink carrier frequency is in the Ku band at 11.9 GHz, the bandwidth block for the UE is 10 MHz, the noise power spectral density is -173 dBm/Hz, the satellite transmit power is 10 dB, the Rician small-scale fading is 20 dB, the atmospheric fading's attenuation is 0.05 db/km. The utility functions of data rate and delay have the same preference, thus, the weights in equation (\ref{eq:weighted-sum-rate-delay}) are both set to 0.5.
% and the weights in \ref{eq:weighted-sum-rate-delay} are both set to 0.5.

The graph-based handover framework is compared against a well-known legacy algorithm for satellite handover handovers. This algorithm is based on a threshold strategy of the satellite's elevation angle from the UE. In this simulation, the threshold angle is 10 degrees. When the satellite elevation angle reaches 10 degrees, the handover process is triggered.

\subsection{Results and Discussion}
In this simulation, data rate and delay are the handover criteria used to customize the handover sequence to the UE. Our graph-based method (GM) and the threshold method (TH) are compared in terms of the resulted UE's data rate. For both methods, the UE starts the connection with the same satellite.

\begin{figure}[htbp]
\centerline{\includegraphics[width=\linewidth]{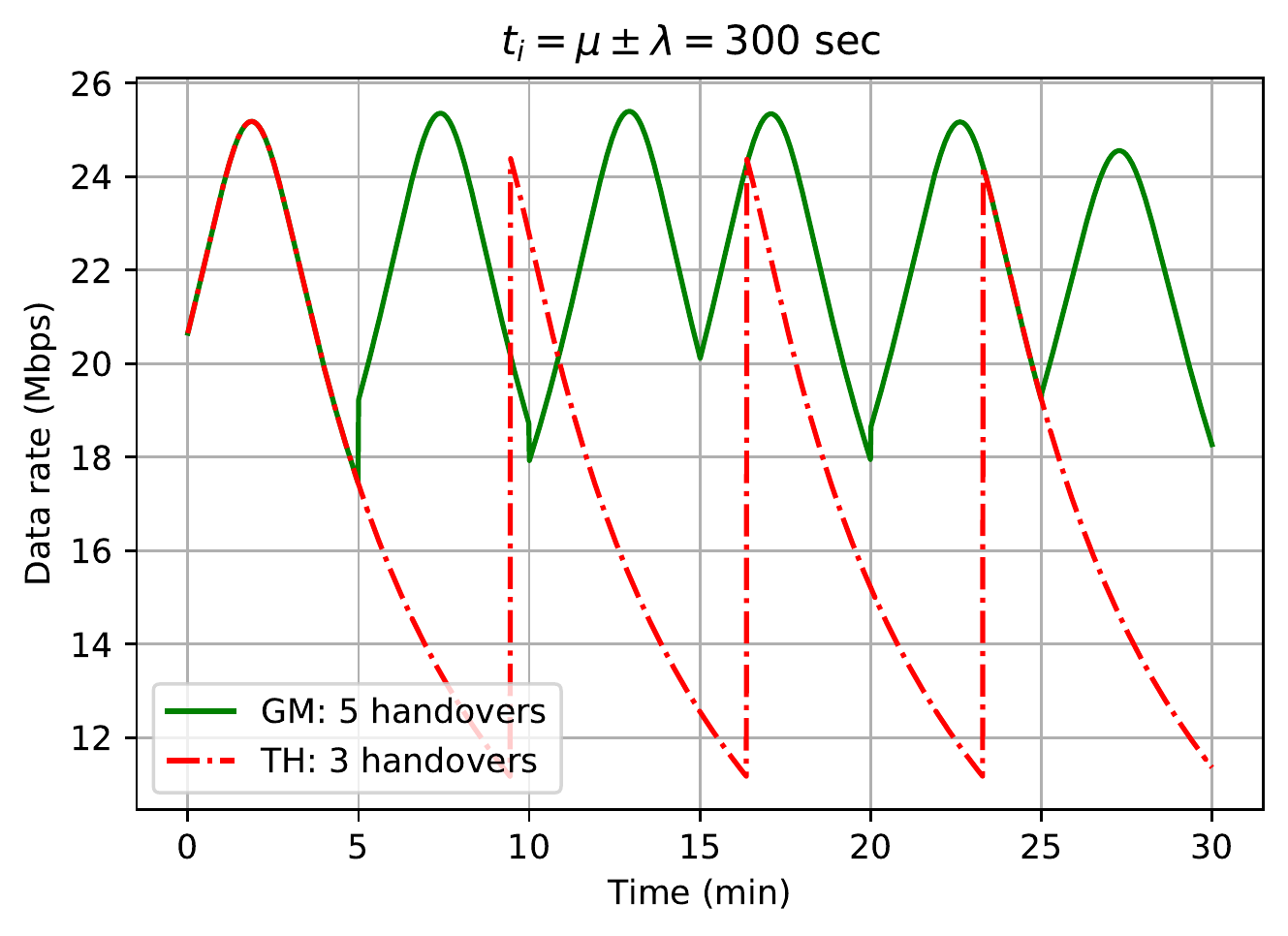}}
\caption{The UE's data rate over 30 minutes using the threshold method (TH) and the graph method (GM) with $\lambda=150$ seconds (2.5 minutes).}
\label{fig:rate-5-min}
\end{figure}

\begin{figure}[htbp]
\centerline{\includegraphics[width=\linewidth]{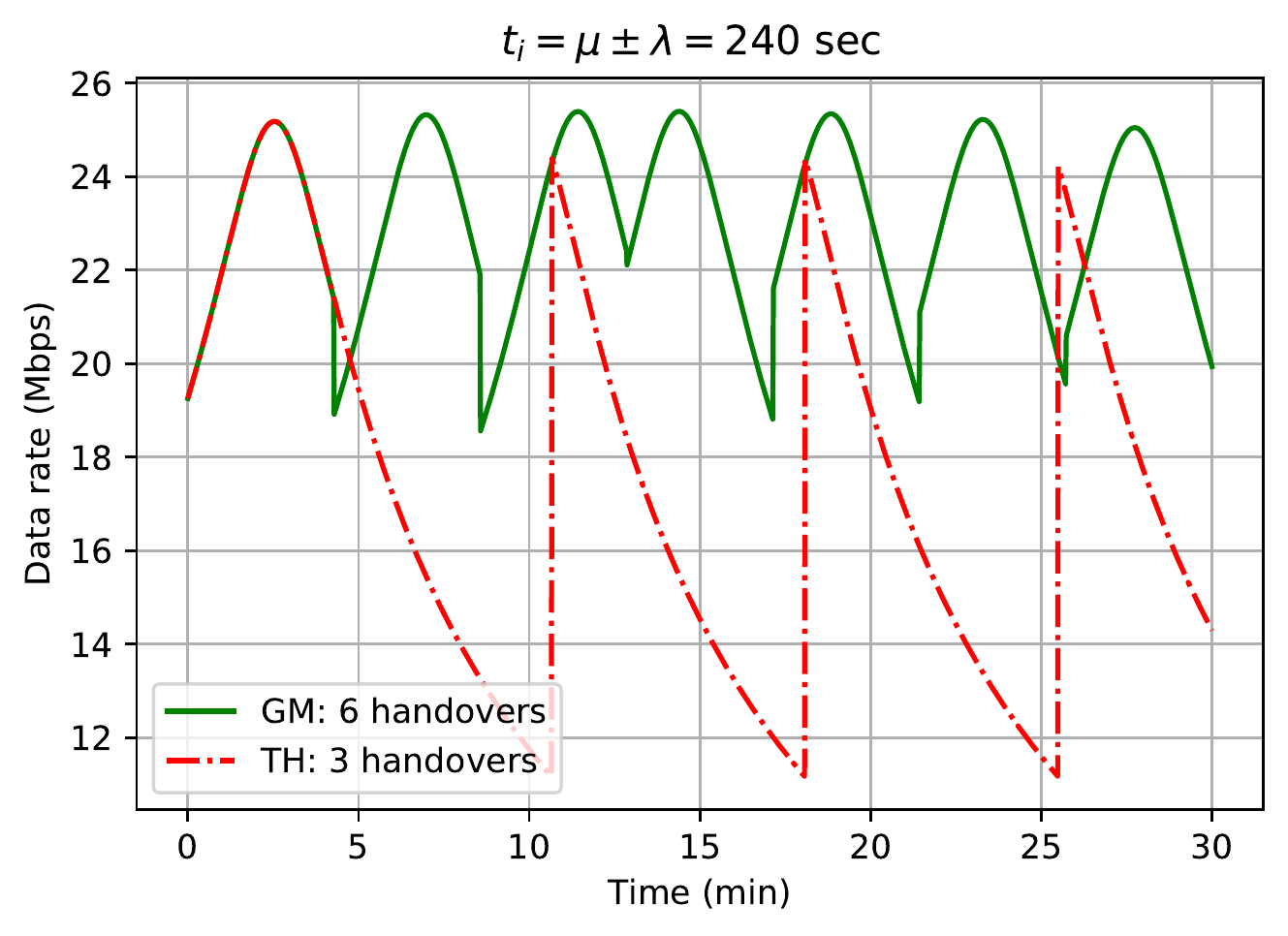}}
\caption{The UE's data rate over 30 minutes using the threshold method (TH) and the graph method (GM) with $\lambda=120$ seconds (2 minutes).}
\label{fig:rate-4-min}
\end{figure}

\begin{figure}[htbp]
\centerline{\includegraphics[width=\linewidth]{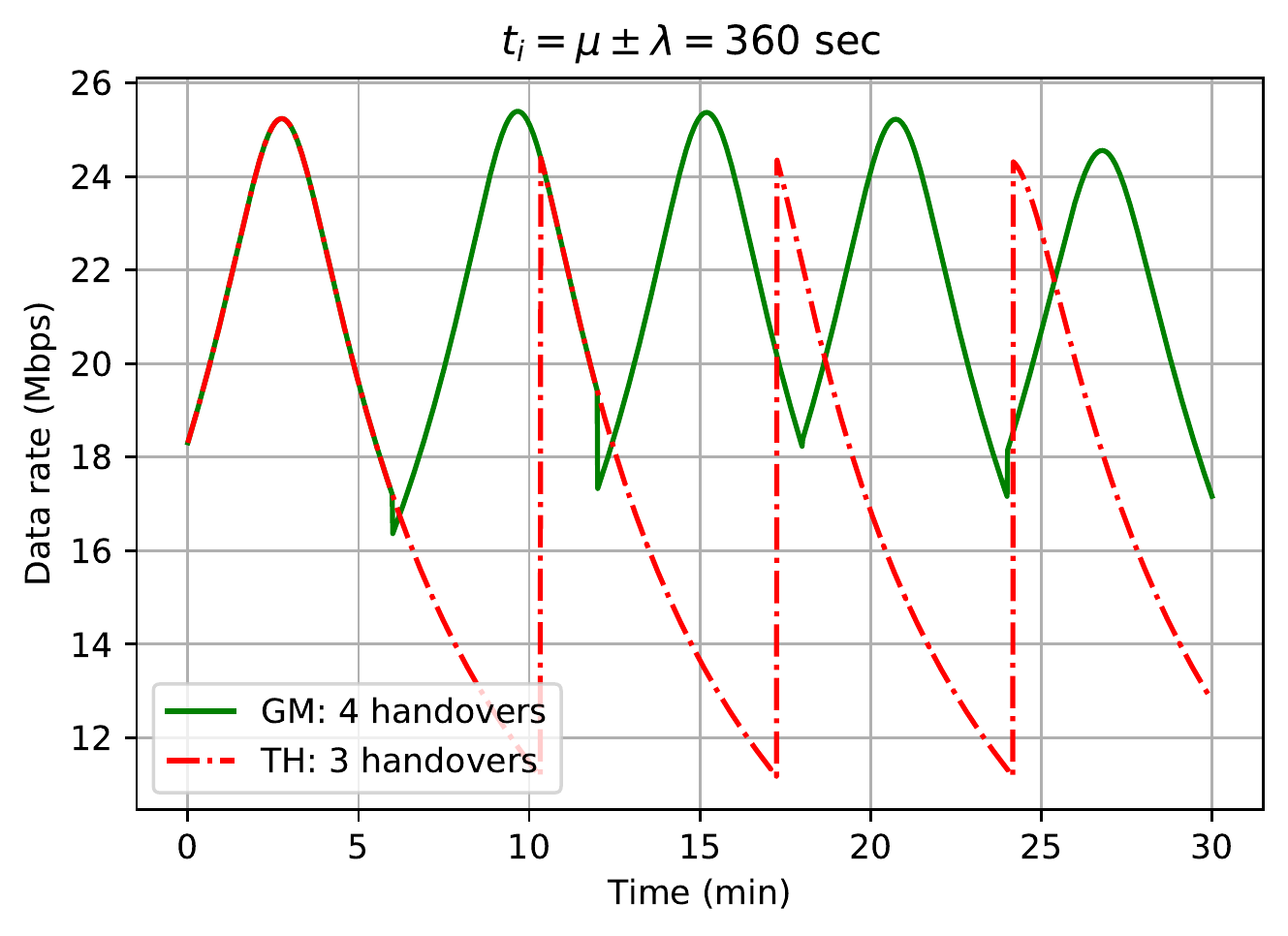}}
\caption{The UE's data rate over 30 minutes using the threshold method (TH) and the graph method (GM) with $\lambda=180$ seconds (3 minutes).}
\label{fig:rate-6-min}
\end{figure}

In Fig. \ref{fig:rate-5-min}, following the customized handover sequence selection process of the GM, the total time duration, $T$, is 30 minutes, which is split into six equally sized time durations, $t_i$. Each time duration is 300 seconds, where $\lambda=150$ seconds, $t_i=(\mu \pm 150)$. During $T$, the TH results in three handovers while the GM results in five handovers at which a handover takes place once every five minutes or every $2 \lambda$. In fact, the GM's number of handovers is dependent on $\lambda$; smaller $\lambda$ values result in suggesting more handovers and vice-versa, as shown in Figs. \ref{fig:rate-4-min} and \ref{fig:rate-6-min}. Although the number of handover suggestions might increase using the GM, with better timing and selection of handovers, soft handovers can be achieved. This is different from the TH method, where the handovers will mostly be hard handovers.

% \begin{figure}[htbp]
% \centerline{\includegraphics[width=\linewidth]{figures/compare-delay-5-min.pdf}}
% \caption{The UE's propagation delay over 30 minutes using the threshold method (TH) and the graph method (GM) with $\lambda=300$ seconds (2.5 minutes).}
% \label{fig:delay-5-min}
% \end{figure}

% \begin{figure}[htbp]
% \centerline{\includegraphics[width=\linewidth]{figures/compare-delay-4-min.pdf}}
% \caption{The UE's propagation delay over 30 minutes using the threshold method (TH) and the graph method (GM) with $\lambda=120$ seconds (2 minutes).}
% \label{fig:delay-4-min}
% \end{figure}

% \begin{figure}[htbp]
% \centerline{\includegraphics[width=\linewidth]{figures/compare-delay-6-min.pdf}}
% \caption{The UE's propagation delay over 30 minutes using the threshold method (TH) and the graph method (GM) with $\lambda=180$ seconds (3 minutes).}
% \label{fig:delay-6-min}
% \end{figure}

\begin{figure}[htbp]
\centerline{\includegraphics[width=\linewidth]{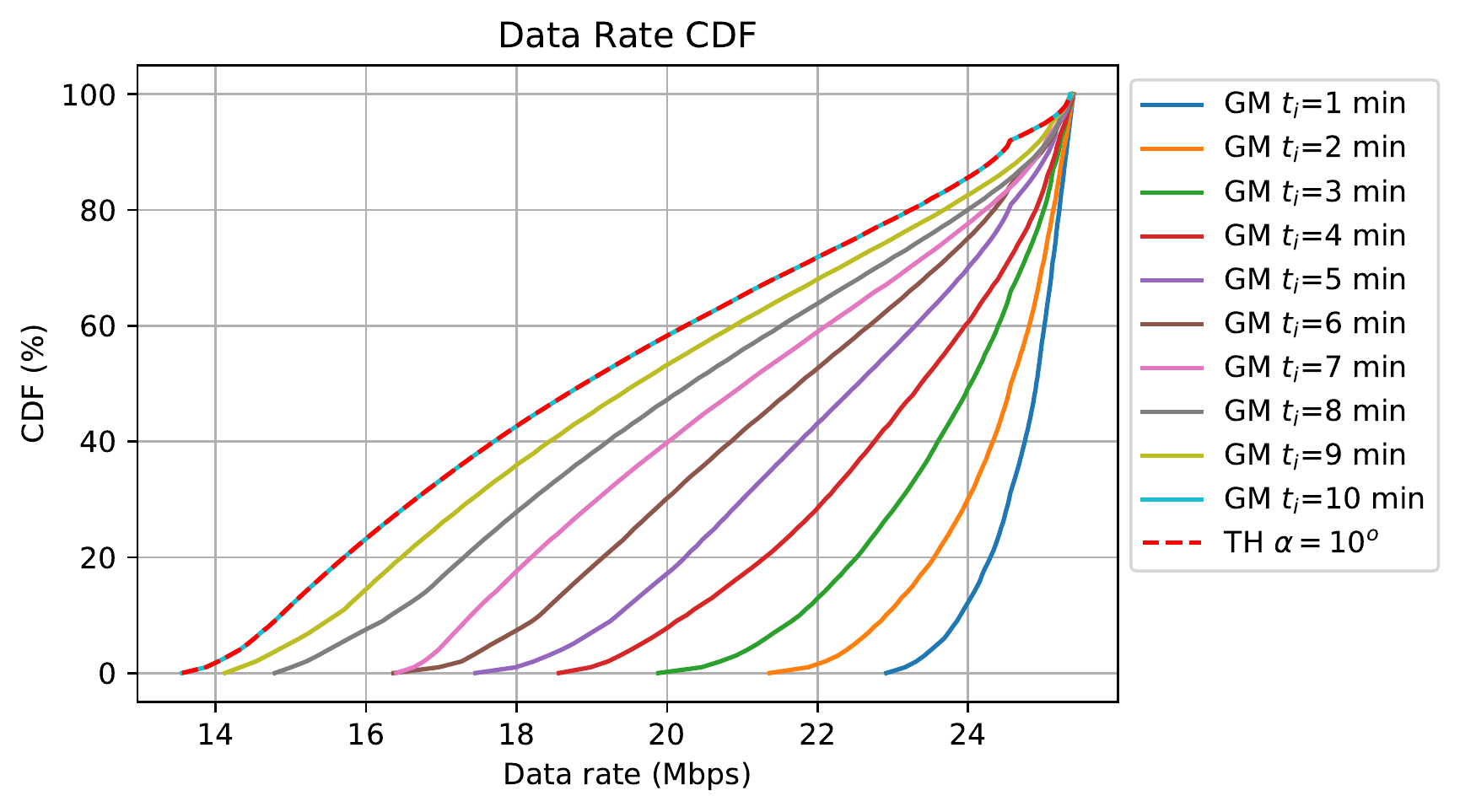}}
\caption{The UE's data rate cumulative distribution function of the threshold method (TH) and the graph method (GM) with various $\lambda$ values ($\lambda = t_i/2$).}
\label{fig:cdf}
\end{figure}

Regarding the performance comparison, by visual inspection, the GM data rate is constrained at a higher range of Mbps than the TH, as shown in Figs. \ref{fig:rate-5-min}, \ref{fig:rate-4-min}, and \ref{fig:rate-6-min}. 
% \ref{fig:delay-4-min} \ref{fig:delay-5-min} \ref{fig:delay-6-min} 
In addition, it is obvious that GM prevents the sharp drops in data rate peaks, which clearly shows that GM has the advantage of maintaining QoS. In particular, the performance of GM in Fig. \ref{fig:rate-5-min}, where GM results in five handovers at which a handover takes place once every five minutes, is perfectly reasonable and much better than in normal terrestrial cellular networks with moving user terminals. Furthermore, a comprehensive statistical comparison is required to quantify the significance of the framework over TH (the legacy algorithm). The cumulative distribution function (CDF) is used to compare the TH and the GM with different $\lambda$ values. Fig. \ref{fig:cdf} shows the data rate behavior of both methods at different CDF percentages. For example, at 20\% CDF, the TH has 20\% of its data rate performance under 15.7 Mbps, which is similar to the GM with $\lambda=300$ seconds (GM-10), GM-6 is under 19.3 Mbps, GM-5 is under 20.1 Mbps, and GM-4 is under 21.2 Mbps. Therefore, as $\lambda$ decreases in the GM, the performance improves while the cost to pay is the number of handovers. A direction for future research would be to analyze the trade off on selecting the appropriate $\lambda$ value considering soft handover and other optimization techniques.

\section{Conclusions}
This paper proposed a graph-based customizable handover planning framework for LEO satellite networks. As discussed, the framework uses the idea of a time-based graph where the satellites instances are represented by vertices and the edge weights are obtained through the customizable handover criteria. The results and accompanying discussion demonstrated shows the effectiveness of the novel handover framework, which is also low in complexity, modular, flexible, and expected to be forward compatible.

\section{Acknowledgment}
This work has been supported by the National Research Council Canada’s (NRC) High Throughput Secure Networks program (CSTIP Grant \#CH-HTSN-625) within the Optical Satellite Communications Consortium Canada (OSC) framework. The authors would like to thank Dr. Aizaz Chaudhry for providing the simulation data of Starlink Phase I constellation. 

\bibliography{BibTex}
\bibliographystyle{IEEEtran}

\end{document}